# A Comparative Study on the Impact of Traditional Learning and Interactive Learning on Students' Academic Performance and Emotional Well-Being

Siva Raja Sindiramutty

Taylor's University, School of Computer Science, Subang Jaya, Malaysia

siva.sindiramutty@taylors.edu.my,

**Abstract**

The growing adoption of interactive learning tools in higher education offers new opportunities to enhance student performance and well-being. This study compares the effects of traditional and interactive learning methods on academic performance, engagement, motivation, and emotional well-being among 100 university students enrolled in a computer intrusion detection course. Participants were randomly assigned to either a traditional learning group (lectures and notes) or an interactive learning group utilising tools such as Kahoot, Panopto, Slido, Quizizz, Padlet, and educational videos. Academic achievement was measured through pre-tests, post-tests, final exams, and assignments, while engagement and emotional states were assessed using validated Likert-scale questionnaires. Results showed that students in the interactive group significantly outperformed their peers in both post-tests (67.48% vs. 53.36%) and final exams (80.8% vs. 61.44%). Interactive learners also demonstrated greater behavioural (+67.01%) and emotional engagement (+75.32%), along with enhanced emotional well-being marked by increased positive emotions (+66.67%) and reduced frustration. A significant drop in cognitive involvement (-39.8%) indicates possible cognitive overload. The pedagogical potential of interactive learning is reaffirmed by this result while reinforcing the need for balancing stimulation and cognitive level. Future research with larger, diverse samples is suggested for generalising and maximising outcomes.

# List of Abbreviations

**MCQ** – Multiple Choice Questions

**PANAS** – Positive and Negative Affect Schedule

**SPSS** – Statistical Package for the Social Sciences

# 1. Introduction

The incorporation of technology in education has transformed instruction, seeking to improve student engagement, motivation, and learning (Schindler et al., 2017). For hundreds of years, traditional education has been marked as an activity that involves teacher-student interaction consisting of organised delivery and evaluation of knowledge (Recker et al., 2024; Barua & Lockee, 2024). But over the past few decades, a major revolution has occurred as interactive learning experiences have evolved to encourage active participation, collaboration, and deeper cognitive involvement (Broadbent et al., 2023). The type of learning that is traditionally known as 'conventional teaching' involves many types of instruction, such as lectures, reading, classroom discussions and exams. Such traditional methods, which are, perhaps, often criticised for supporting passive learning, are firmly grounded in pedagogic theories (behaviourism, cognitivism) and may include active approaches (Socratic questioning, case-based learning) when enthusiastically implemented (Pritchard & Woollard, 2013; Mangtani, 2024). Hence, traditional learning is not a priori without interactivity or learner-focused sensibilities.

Interactive learning is described as a development in pedagogical practices toward student-centred models, and it focuses on participation, critical thinking, cooperation, and the employment of technology-based tools (Suárez & El-Henawy, 2023). Kahoot, Quizizz, Padlet, and Panopto, among others, offer opportunities for game-based learning, digital learning environments with immediate feedback, and types of experimental learning activities that are consistent with constructivist and experiential learning theories (Jonassen & Rohrer-Murphy, 1999; Piaget, 1954; Vygotsky, 1980). Nevertheless, interactive learning is not confined to the cyber-world but encompasses all teaching methods aiming to engage students through discussions, peer teaching and group activities. Research has found that interactive learning environments lead to increased levels of student motivation, engagement, emotional well-being, and knowledge retention than do more traditional lecture-based educational environments (Khozaei et al., 2022; Capone & Lepore, 2021; Song & Cai, 2024). However, issues such as cognitive overload, digital fatigue, and technological barriers could potentially hinder the effectiveness of these strategies if not well conceived (Skulmowski, 2024; Fasoli, 2021). Because of increasing focus on educational innovation, an assessment of the relative effectiveness of traditional and modern teaching approaches is required. A consideration of the advantages and disadvantages of such approaches can inform pedagogical decisions and ultimately may lead to the construction of more effective, engaging, and inclusive educational environments.

This research aims to explore how traditional versus interactive learning mode influences university students' academic achievement and engagement, motivation, and emotional contentment in a computer intrusion detection discipline. Overall, by using an experimental design and a mixed-methods approach, this study hopes to provide empirical findings for evidence-based best practices in contemporary education.

## 2. Literature Review

### 2.1 Traditional Learning: Foundations and Pedagogical Relevance

Traditional learning refers to long-established educational methods such as lectures, textbook reading, in-class discussions, and structured assessments. While often critiqued for promoting passive learning, these approaches are rooted in pedagogical theories such as behaviorism and cognitivism, which emphasize direct instruction, repetition, and hierarchical knowledge transfer (Pritchard & Woollard, 2013; Kong & Wang, 2024). When effectively implemented, traditional teaching can also foster active engagement through methods like case studies, Socratic questioning, and collaborative debates, especially in theoretical and writing-intensive disciplines.

However, limitations exist. Research has shown that traditional methods may be less engaging for digital-native learners and may fail to support differentiated instruction or collaboration (Shloul et al., 2024). Critics argue that reliance on lectures can lead to surface-level learning, limited interaction, and decreased student motivation (Lawter & Garnjost, 2021).

### 2.2 Interactive Learning: Pedagogical Principles and Tools

Interactive learning is characterised by active student participation, immediate feedback, peer collaboration, and problem-solving. It aligns strongly with constructivist theory (Piaget, 1954; Vygotsky, 1980) and experiential learning (Kolb, 1984), which advocate for learners to construct knowledge through engagement and experience.

Digital tools such as Kahoot, Slido, Padlet, and Quizizz have enabled widespread implementation of interactive strategies in higher education (Suárez & El-Henawy, 2023). These platforms enhance gamification, real-time feedback, and multimodal content delivery — all of which contribute to increased engagement and motivation (Capone & Lepore, 2021; Liu et al., 2024). However, interactive learning is not exclusively digital; non-technological techniques such as think-pair-share, role-play, and peer-led tutorials also qualify under this framework.

Despite its advantages, interactive learning may lead to cognitive overload, especially when poorly structured or excessively stimulating (Skulmowski, 2024). Digital fatigue, distractions, and inequalities in access to devices and internet connectivity also pose barriers to effective implementation (Fasoli, 2021; Nobles, 2022).

## 2.3 Student Engagement and Motivation

Student engagement — cognitive, emotional, and behavioural — is essential for meaningful learning. Comparative research indicates that interactive learning environments tend to produce higher levels of engagement through gamification and collaborative activities (Chans & Castro, 2021). Students often report greater enjoyment, enthusiasm, and self-direction in interactive settings compared to traditional lecture-based models (Reyes et al., 2012; Tzafilkou et al., 2021).

However, traditional learning provides predictability, structure, and clear performance expectations, which benefit some learners, particularly those who struggle with self-regulation in highly autonomous environments (Pourabedin & Biglari, 2024). Overreliance on interactivity without reflective design may lead to superficial engagement, focused more on fun than depth.

## 2.4 Academic Achievement and Cognitive Development

Several investigations demonstrate that interactivity and technology-enhanced learning are beneficial for enhancing academic learning, particularly in the field of STEM, where simulation and gamified practice games facilitate the understanding of concepts (Zhao et al., 2022; Huang et al., 2022). Both virtual labs as well as scenario activities are particularly suitable for education in medicine and engineering, which can provide safe training environments for complex skills (Lozano-Durán et al., 2023).

However, old-style learning is not completely without value when it comes to a subject which demands deep theoretical thought, such as philosophy, literature, or law. Organised reading, writing and instructor-facilitated discussions help students develop analytical and critical thinking skills in a structured way (Chen, 2020; Broadbent et al., 2023). This indicates the importance of subject-specific instructional design when comparing learning approaches.

## 2.5 Emotional Well-being and Cognitive Load

Mental health is tied to how well students do in school. Interactive environments might motivate positive affective states (e.g., enjoyment, curiosity, and low anxiety) because of their interactive and encouraging nature (Zhou, 2025; Wang et al., 2022). Nevertheless, research has suggested that heavy multimedia stimuli could cause cognitive load, mental fatigue, and decreased depth of processing (Bernhardt & Poltavski, 2020; Skulmowski & Xu, 2021). In contrast, face-to-face classrooms provide emotional security and minimise distractions that are helpful for those students who have a preference for predictable and low-stimulating settings (Kruk et al., 2022). Thus, emotional and psychological outcomes differ by student preferences, the quality of the instruction, and the nature of the instruction

## 2.6 Summary and Theoretical Implications

There are pros and cons to both traditional teaching and interactive teaching. However, rather than considering them as competing perspectives, a hybrid approach, based on constructivist, cognitive load and self-determination theories, might be the most pedagogically appropriate (Zhou et al., 2023) This review highlights that instructional design should be based on an evidence-base of effective strategies and adapted to the learner, content, and mode of delivery. It also underscores the need to measure not only performance but also engagement, emotional well-being and cognitive effort — dimensions that this study seeks to investigate in a controlled university environment. Table 1 shows a pedagogical and practical comparison between traditional and interactive learning approaches.

Table 1: Pedagogical and Practical Comparison Between Traditional and Interactive Learning Approaches

| Aspect | Traditional Learning | Interactive Learning |
|---|---|---|
| **Pedagogical Basis** | Behaviourism, Cognitivism | Constructivism, Experiential Learning |
| **Aspect** | Traditional Learning | Interactive Learning |
| **Strengths** | Structured and predictable, Effective for deep reading/writing tasks, Clear assessment standards, Lower cognitive load. | Enhances engagement and motivation, Promotes collaboration and active learning, Immediate feedback, Stimulates curiosity and enjoyment. |
| **Limitations** | It can promote passive learning, is less responsive to diverse learning styles, and may not engage digital-native learners. | Risk of cognitive overload, Digital distractions, requires a strong digital infrastructure. |
| **Best Suited For** | Theoretical disciplines (law, philosophy), Students needing structure, In-depth critical thinking. | STEM, skill-based learning, Project-based or collaborative work, Digital-native students |
| **Emotional Impact** | Stable, low-stimulation; may be boring or disengaging | High stimulation, emotionally engaging, can cause fatigue or overwhelm. |
| **Assessment Style** | Exams, essays, and structured assignments | Quizzes, polls, simulations, peer evaluation |

## 3. Methodology

### 3.1 Research Design

This study adopted a true experimental approach, with a between-subjects design. Participants were randomly divided into two groups (one with conventional learning and one with interactive learning tools). The research followed a pre-test/post-test/final test design that included measures of academic achievement and emotional/engagement variables. Both quantitative and qualitative data were collected, utilising a mixed-methods strategy to facilitate data triangulation and to permit a deeper understanding.

## 3.2 Participants and Sampling

The study was conducted among 100 Malaysian university undergraduate students who were enrolled in a computer science program. Participants were aged between 18 and 25 years and were recruited via an open call to a first-year computing class. Sampling was stratified by gender and self-reported digital literacy, to enhance diversity and reduce bias. After stratification, participants were randomly allocated into two equal groups, Traditional Learning Group (n = 50) and Interactive Learning Group (n = 50) with All students provided informed consent.

## 3.3 Learning Intervention

The experimental intervention was conducted over fourteen weeks and consisted of twelve instructional sessions, each lasting one hour. Each of the Traditional and Interactive Learning Groups addressed the identical twelve core subjects concerning computer intrusion detection, characterising the course materials as equivalent across conditions. All training sessions were conducted by the same researcher to ensure instructional consistency. In the Traditional Learning Group, students engaged in didactic, lecture-based learning, which focused on textbook readings, PowerPoint presentations, and structured in-class discussions. Study materials were in print and evaluations were paper-based.

The Interactive Learning Group, on the other hand, were taught using a range of digital learning tools, such as Kahoot, Quizizz, Padlet, Panapto and Slido. These applications were employed to deliver course content, to engage in active learning exercises, and to administer instantaneous quizzes and feedback. The interactive features were situated within a pedagogical context that drew on constructivism and the principles of experiential learning. Students interacted with multimedia presentations, polls, and games and worked together on assignments in a technology-enhanced setting. Both sets followed the same schedule of topics, methods of assessment, and learning outcomes.

## 3.4 Measurement Instruments

### 3.4.1 Academic Achievement

Academic performance was evaluated by pre-test, post-test and final examination. The pre-test was given before conducting the intervention, and it comprised 30 multiple-choice questions (MCQs) that targeted the first six subject matters of the course. After the sixth session, participants completed a post-test with 30 MCQs on the same content that had been taught. This protocol enabled us to assess short-term learning gains. The MCQ test was given as a final, which was conducted at the end of the course included 60 items covering the entire twelve topics. To ensure content validity and correspondence with learning objectives, all test questions were scrutinised by three experts in the field. The format, difficulty-level, and timing of the tests were made comparable for both groups. Item reliability estimates were not computed in this study, but question development was consistent with traditional instructional design and test construction principles.

### 3.4.2 Student Engagement and Motivation

Student engagement was measured using a 15-item survey adapted from the framework developed by Reyes et al. (2012), which includes behavioural, emotional, and cognitive engagement dimensions. Students responded using a five-point Likert scale ranging from 1 (strongly disagree) to 5 (strongly agree). Academic motivation was assessed using a 9-item scale adapted from Tzafilkou et al. (2021), measuring intrinsic motivation, extrinsic motivation, and amotivation, using the same five-point Likert format.

The original versions of these instruments have demonstrated strong internal consistency in prior research, with reported Cronbach's alpha values of 0.79 for the engagement scale and 0.84 for the motivation scale. In this study, the items were used without major modification, and while reliability was not recalculated on the study sample, the cited reliability values support the instruments' internal consistency.

### 3.4.3 Emotional Well-being

Emotional state was measured with an affect scale based on items from the Positive and Negative Affect Schedule (PANAS) tailored to address students' emotional experiences during learning. The measurement device involved markers for positive emotions (enjoyment, enthusiasm) and negative affect states (frustration, anxiety, boredom). Some other measures were also added to assess students' perceived stress measures as well as general affective loading throughout the study. The PANAS instrument has been widely validated in academic research, with Cronbach's alpha values typically above 0.80 reported. The version used in this study was adapted for the academic context with minimal changes, and its use was intended to preserve its validated structure and psychometric reliability.

### 3.5 Data Collection Procedures

The data collection process was structured to align with the instructional timeline. In Week 1, students completed the pre-test before the start of the intervention. The instructional sessions spanned Weeks 2 through 6, during which both groups received their respective treatments. The post-test was administered at the end of Week 6 to measure learning gains following the initial half of the course content. The final exam was conducted in Week 7 to assess cumulative learning outcomes.

Survey instruments measuring engagement, motivation, and emotional well-being were distributed to students in both groups after the intervention period. All surveys were completed anonymously, and participation was voluntary. In total, three phases of data collection occurred: baseline (pre-test), mid-intervention (post-test), and post-intervention (final exam and surveys).

### 3.6 Data Analysis

Quantitative data were analysed using SPSS version 26. Descriptive statistics, including means and standard deviations, were calculated to summarise students' performance and affective outcomes. Inferential statistical tests were applied to determine the significance of differences between groups.

To compare academic performance outcomes while accounting for initial ability, analysis of covariance (ANCOVA) was used, with the pre-test score serving as the covariate and the post-test and final exam scores as dependent variables. Independent samples t-tests were employed to analyse differences between the two groups in engagement, motivation, and emotional well-being scores. Cohen's d was calculated to assess effect sizes and interpret the practical significance of group differences. Statistical significance was set at $p < 0.05$. All data were checked for normality, and assumptions of homogeneity of variance were verified before conducting the inferential analyses.

### 3.7 Ethical Considerations

Students were informed of the study's purpose, data confidentiality, and their right to withdraw at any time without penalty.

## 4. Results and Discussion

This section presents an in-depth interpretation of data gathered in this pilot study on students' learning, affect states, motivation, engagement, and acceptance of technology. The current literature is integrated to draw salient insights and future directions for metaverse-enabled learning.

## 4.1 Academic Performance

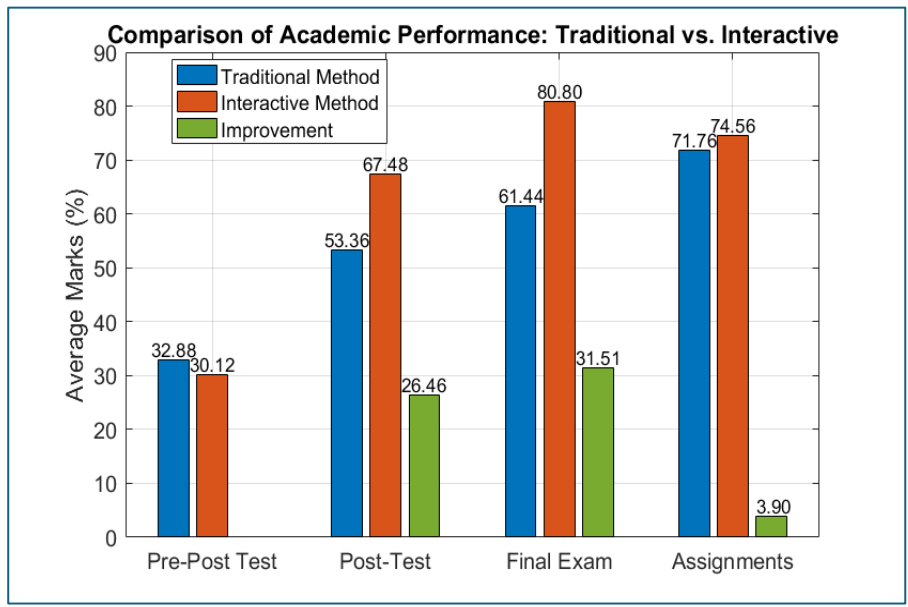

Figure 1: **Comparison** of academic performance

Figure 1 presents the comparative analysis of academic performance between students in the traditional and interactive learning groups. To assess the impact of instructional method on student learning outcomes, ANCOVA was used to compare post-test and final examination scores between the traditional and interactive learning groups, using pre-test scores as a covariate. This controls for initial knowledge differences and isolates the effect of the instructional method. The **pre-test results** indicated that students in both groups had comparable baseline knowledge. Traditional learners scored an average of **32.88%**, while interactive learners scored **30.12%**, ensuring that observed differences are not attributable to pre-existing disparities in competency.

After adjusting for baseline knowledge, a significant difference was found in **post-test scores**, **F(1,47) = 5.19, p = .027**. The interactive learning group achieved a mean score of **67.48%**, compared to **53.36%** in the traditional group — a **26.46% greater improvement**, suggesting that gamified learning enhances short-term retention and conceptual understanding. This aligns with prior research on digital learning environments that emphasise active participation and contextual engagement (**Ben**, 2025).

A more pronounced effect was observed in the **final examination scores**, covering all ten course topics. The ANCOVA revealed a highly significant group difference, **F(1,47) = 36.87, p < .001**, with the interactive group scoring **80.80%** on average, far surpassing the traditional group's **61.44%** — a **31.51% improvement**. This finding reinforces the long-term benefit of interactive instruction for deeper learning and application, in line with theories emphasising knowledge transfer and higher-order thinking (Maceiras et al., 2025).

Regarding **assignment performance**, students in the interactive group performed slightly better (**74.56%**) than those in the traditional group (**71.76%**). However, the **3.90% difference** was marginal, suggesting that interactive tools may not substantially improve tasks involving written articulation and extended reasoning. Nevertheless, scenario-based interactivity may still contribute to enhanced analytical skills in applied or decision-making contexts.

### 4.2 Emotional Well-Being and Engagement

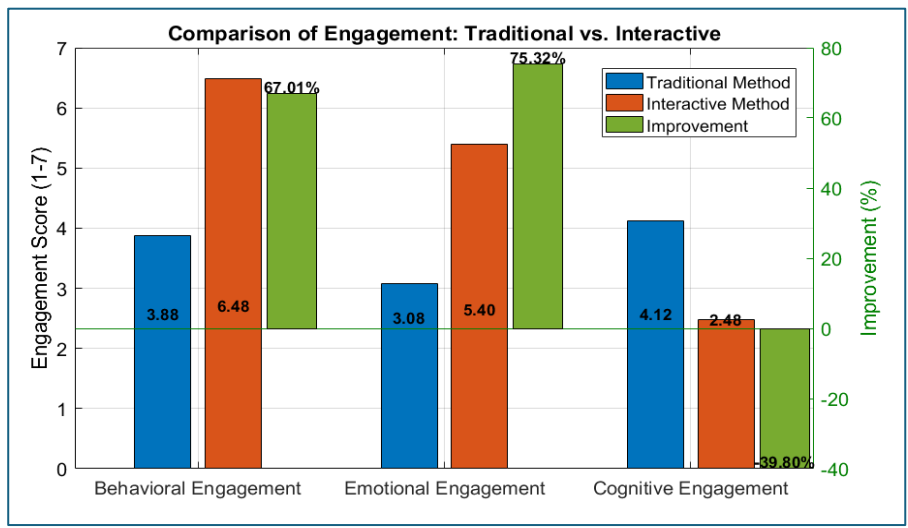

Figure 2: Comparison of academic engagement

Figure 2 illustrates student engagement levels across both learning environments. Student engagement was assessed across three dimensions: behavioural, emotional, and cognitive. Independent samples t-tests revealed statistically significant differences favouring the interactive learning group in behavioural and emotional engagement, while cognitive engagement was higher in the traditional group.

**Behavioural engagement** was significantly greater among interactive learners (**M = 6.48**) compared to traditional learners (**M = 3.88**), **t(48) = 5.70, p < .001**, with a **very large effect size (Cohen's d = 1.61)**. This 67.01% increase reflects students' active involvement in classroom activities, likely driven by real-time interaction, gamified elements, and feedback loops available in tools like Kahoot and Padlet.

Similarly, **emotional engagement** was notably higher in the interactive group (**M = 5.40**) than the traditional group (**M = 3.08**), **t(48) = 4.55, p < .001, d = 1.29**. This 75.32% gain aligns with prior research indicating that dynamic, participatory environments foster stronger affective bonds with learning material (Tzafilkou et al., 2021).

However, the results revealed a reverse trend in **cognitive engagement**, with traditional learners scoring significantly higher (**M = 4.12**) than their interactive peers (**M = 2.48**), **t(48) = -3.72, p = .001, d = -1.05**. This 39.81% decline suggests that interactive methods, while emotionally and behaviorally stimulating, may overwhelm students cognitively. This supports earlier findings on **cognitive overload**, where excessive stimuli hinder deep processing and critical reflection (Skulmowski, 2024; Bernhardt & Poltavski, 2020).

These findings highlight the need to balance engagement strategies, promoting activity and excitement without compromising reflective thinking.

### 4.3 Student Motivation

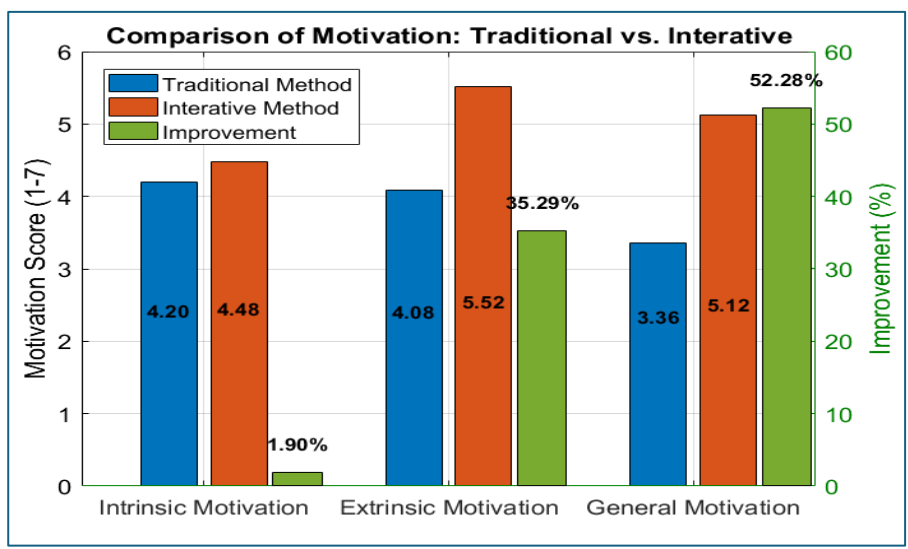

Figure 3: Comparison of academic motivation

Figure 3 presents students' motivation levels across different dimensions. Motivation was analysed through intrinsic, extrinsic, and general (combined) dimensions. The interactive group scored significantly higher in **extrinsic** and **general motivation**, though **intrinsic motivation** differences were not statistically significant.

**Extrinsic motivation** was significantly higher in the interactive group (**M = 5.52**) than in the traditional group (**M = 4.08**), **t(48) = 3.85, p = .001, d = 1.09**, indicating a 35.29% increase. This suggests that students responded positively to external cues such as competition, instant rewards, and social feedback — features commonly embedded in gamified learning environments.

**General motivation**, a combined index of motivational drivers, also favoured the interactive group (**M = 5.12**) over the traditional group (**M = 3.36**), **t(48) = 3.75, p = .001, d = 1.06**, reflecting a 52.38% increase. This suggests that students in interactive environments feel more energised and goal-directed overall, potentially due to the autonomy and variety in activities.

However, **intrinsic motivation** — the internal desire to learn for personal satisfaction — showed no significant difference (**M = 4.48** vs. **4.20**, **p = .590**). This aligns with self-determination theory (Ryan & Deci, 2000), which posits that intrinsic motivation is harder to influence externally and may require sustained intellectual curiosity, not just fun or novelty.

These results suggest that interactive tools effectively engage students through structured rewards and variety but may not automatically foster deeper, self-driven interest in the subject unless pedagogically scaffolded.

## 4.4 Emotional States & Well-being

Figure 4 provides insights into students' emotional responses during learning. Emotional well-being was assessed through three components: positive emotion, negative emotion, and flow state. Students in the interactive learning group reported significantly more positive emotional experiences, lower negative emotions, and greater flow — the immersive state of focused engagement.

**Positive emotional states**, such as enjoyment and enthusiasm, were higher in the interactive group (**M = 4.00**) compared to the traditional group (**M = 2.40**), **t(48) = 5.66, p < .001, d = 1.60**, representing a 66.67% increase. This suggests that interactive methods help reduce boredom and promote engagement, consistent with findings from Zhou (2025) and Wang et al. (2022).

Conversely, **negative emotions** such as frustration, anxiety, and confusion were significantly lower in the interactive group (**M = 2.56**) than in the traditional group (**M = 4.00**), **t(48) = -4.88, p < .001, d = -1.38**, a 36.00% reduction. This supports the argument that digital environments, when well-designed, can reduce performance pressure by offering safe, trial-and-error-based learning.

**Flow state**, reflecting immersion and psychological engagement, was also higher in the interactive group (**M = 4.00**) than in the traditional group (**M = 2.44**), **t(48) = 5.62, p < .001, d = 1.59**. This 38.54% improvement underscores the role of technology-enhanced instruction in promoting sustained attention and satisfaction during learning tasks.

Collectively, these findings suggest that interactive learning environments not only engage students externally but also support their emotional resilience and psychological involvement, key factors in long-term academic persistence and satisfaction.

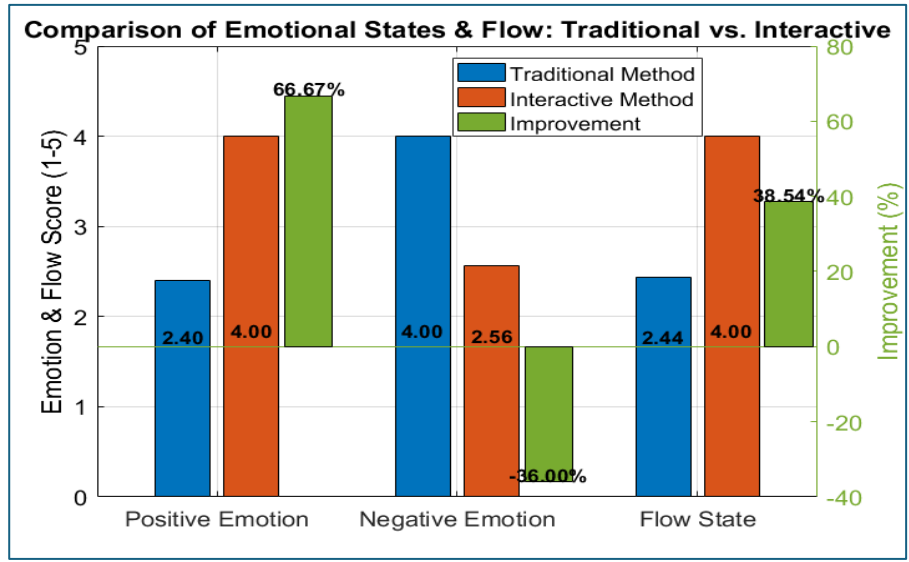

Figure 4: Comparison of emotional and flow

## 4.5 Measuring Technology Acceptance

Figure 5 evaluates students' perceptions of the interactive learning platforms. The perceived ease of use was rated at 5.24, indicating that students found the platforms relatively intuitive and navigable. Perceived usefulness, measuring the extent to which students felt that interactive learning enhanced their understanding, scored 5.4, demonstrating that students acknowledged the benefits of the technology. Additionally, satisfaction and engagement were notably high, with students scoring 6.0, suggesting a preference for interactive learning over traditional methods. These findings align with studies that emphasise usability and engagement as critical factors in technology adoption for education (Kim et al., 2025)

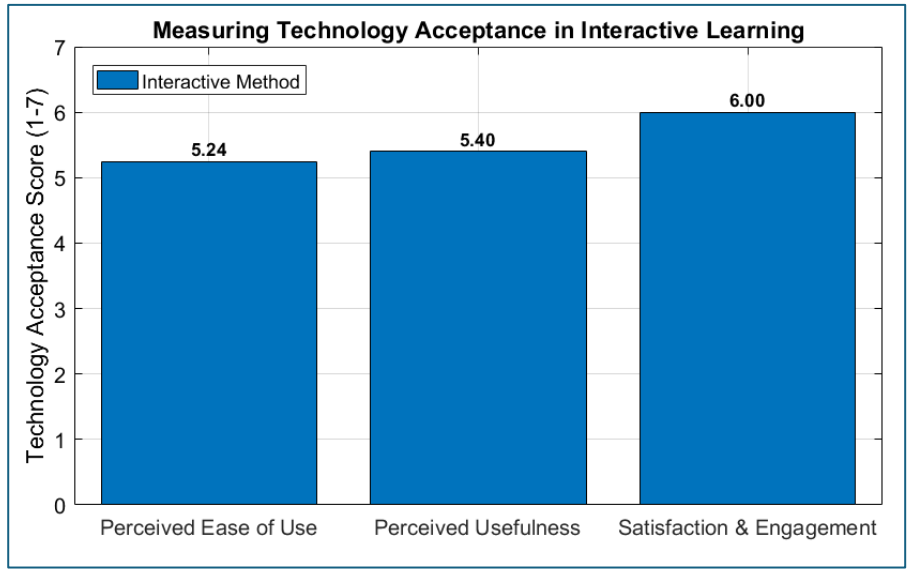

Figure 5: Measuring technology acceptance

## 4.6 Discussion and Implications

This study highlights several areas for future research to optimise interactive learning's impact and scalability. For one, future studies should identify strategies for interactive learning engagement quality to handle the cognitive load and avoid digital fatigue, ensuring that interactive methods do not overwhelm students or stifle deeper cognitive activity. Longitudinal studies also must take place to assess interactive learning's effects on scholastic application and retention of acquired knowledge in the long run, in case short-run benefits do not convert to lasting scholastic performance. Research needs to develop scalable models for including interactive learning in standard education and education systems, and studies on larger and diverse student bases to assess generalizability and feasibility. Addressing these areas will optimise interactive learning to better maintain students' learning in an efficient and educationally sound manner.

## 4.7 Future Research Directions

The findings of this study provide a comprehensive evaluation of the differential effects of traditional and interactive learning approaches on academic performance, engagement, motivation, and emotional well-being. Consistent with prior literature, the results suggest that **interactive learning environments offer substantial benefits** in promoting behavioural and emotional engagement, extrinsic motivation, and positive affect. However, the **reduction in cognitive** **engagement** among interactive learners raises important concerns about the potential for **cognitive overload** when digital tools are overused or poorly scaffolded.

From a **constructivist perspective**, interactive learning supports active meaning-making, peer collaboration, and feedback-rich tasks, which can explain the significantly higher performance in both the post-test and final examination. The tools used — Kahoot, Quizizz, Padlet, and others — offer gamified and immediate-response environments that reinforce content through repetition, competition, and enjoyment. This aligns with the literature asserting that digital interactivity enhances short- and long-term retention through multisensory learning and contextual reinforcement (Capone & Lepore, 2021; Huang et al., 2022).

The **engagement patterns** further underscore this advantage. Students in the interactive group showed dramatically higher levels of behavioural and emotional engagement, but significantly lower levels of cognitive engagement compared to their traditional counterparts. This suggests that while interactivity increases students' willingness to participate and enjoy the learning process, it may not always translate into deep, reflective processing of material — a problem well-documented in **cognitive load theory** (Sweller, 2011; Skulmowski, 2024). Rapid switching between digital tasks, superficial gamification, or a lack of mental "quiet time" may interfere with conceptual integration and critical thinking.

In terms of **motivation**, the study found that interactive learning environments significantly improved **extrinsic and general motivation**, while having no impact on **intrinsic motivation**. This aligns with **self-determination theory** (Ryan & Deci, 2000), which suggests that external rewards and stimulating conditions can drive participation but do not necessarily foster deep personal interest unless learners are autonomously involved in decision-making and meaning construction.

The emotional well-being results — increased positive affect, lower negative emotion, and higher flow states — highlight a key benefit of digital interactivity: its capacity to humanise the learning experience. When students feel emotionally safe and cognitively immersed, they are more likely to persist, participate, and perform. These affective gains, in turn, reinforce behavioural engagement and motivation, creating a virtuous cycle that supports academic outcomes.

**Implications for practice** include the need for **balanced instructional design**. While interactive tools can greatly enhance motivation and engagement, educators must structure them carefully to avoid overstimulation and ensure cognitive depth. For example, combining gamified quizzes with metacognitive reflection activities may help sustain both enjoyment and conceptual rigour. Blended learning models — mixing traditional explanations with interactive tasks — may offer the optimal strategy to leverage the strengths of both methods.

For **educational researchers**, the study provides empirical evidence that supports a nuanced view of interactivity, not as a universal solution, but as a tool that requires thoughtful integration. Future work should explore adaptive systems that respond to students' cognitive load levels in real time, and comparative studies across disciplines would help determine when and where interactive learning is most effective.

# 5. Conclusion

This study investigated the impact of traditional and interactive learning methods on students' academic performance, engagement, motivation, and emotional well-being within a university-level computing course. Using an experimental design with pre-test, post-test, and final examination data, along with validated engagement and emotion surveys, the study offers robust insights into how instructional methods affect learners across cognitive and affective domains.

The findings demonstrate that **interactive learning environments significantly enhance behavioural and emotional engagement**, **extrinsic motivation**, **positive emotion**, and **academic performance** in both short- and long-term assessments. These outcomes are consistent with constructivist and experiential learning theories, which emphasise the value of student-centred, participatory learning experiences. Interactive tools provided timely feedback, stimulated motivation through gamification, and fostered emotionally engaging classrooms.

However, the study also uncovered important limitations of interactive methods. Students in the interactive group reported **lower levels of cognitive engagement**, suggesting that while these environments may be more stimulating and enjoyable, they can also result in **cognitive overload** when not carefully designed. Furthermore, **intrinsic motivation** — a critical predictor of lifelong learning — remained unaffected, implying that interactive features alone are insufficient to cultivate deep, self-directed interest in subject matter.

In conclusion, interactive learning tools hold substantial promise for improving engagement and performance, but they should be implemented with pedagogical precision. Educators must strike a balance between stimulation and structure to ensure both emotional satisfaction and deep learning. This study contributes to a more nuanced understanding of when and how interactive learning should be employed, and it offers guidance for the next generation of technology-enhanced education.

## 6. Future Recommendations

To further advance interactive learning, future research should expand in several important areas, including breadth of studies, instructional design, user experience, psychological and affective effects, and mixed models of learning. Expanding the range of studies is warranted to promote generalizability. Larger student populations in various institutions, to account for variability in student demographics, learning contexts, and education settings, should be included in subsequent studies. Longitudinal studies, in which data is gathered on several occasions, would better capture long-term retention of learning and cognitive development. Adopting such an approach, scholars would have the ability to assess interactive learning's long-lasting effects on scholastic achievements and psychological well-being. Improvements in instructional design are also necessary to address interactive learning's cognitive engagement problems.

Systematic learning routes have to be developed to balance immersive and reflective learning, where students critically interact with content and do not passively engage with digital tools. Moreover, including inquiry-guided exercises may promote critical thinking by having students approach content through problem-solving and analytical thinking. Interactive evaluation through gamified and problem-solving exercises needs to be integrated to advance student learning and engagement. Optimising the student's experience is equally imperative for future adoption and implementation. Simplification of design in the interface to reduce cognitive load allows students to focus on learning rather than grappling with overly complex digital environments. System and student and instructor training in a structured format is necessary to ensure smooth adoption and enhanced application of interactive tools.

Resolution of technical problems such as lag in connectivity and system stability is equally crucial to maintaining students' attention and uninterrupted learning. Beyond academic achievement and performance, future work has to examine interactive learning's affective and psychological impacts. Longitudinal work has to examine to what extent extensive digital tool use affects pupils' welfare in terms of stress, motivation, and cognitive load. One area has to be determining an appropriate balance between in-depth work and activity to maintain stimulated pupils while not overloading pupils cognitively through the overuse of digital tools. Utilizing hybrid modes of teaching can turn out to be an ideal antidote by combining interactive benefits with traditional benefits. Utilising flipped modes where interactive tools are applied to practice classes and theoretical classes in classrooms can turn out to have an overall boost in learning. Furthermore, personalised pathways to allow pupils to alternate between interactive modes and traditional modes on individual needs can turn out to be more responsive and adaptable in teaching.

Summarily, through this work, interactive learning is revealed to have vast potential to drive academic achievement, student engagement, and student welfare. Research does point to areas where work is necessary, including in instructional design and cognition. Future work should be to optimize interactive teaching to achieve scalability, sustainability, and effectiveness in future tertiary teaching.

Zhao, D., Muntean, C. H., Chis, A. E., Rozinaj, G., & Muntean, G. (2022). Game-Based Learning: Enhancing student experience, knowledge gain, and usability in higher education programming courses. *IEEE Transactions on Education*, *65*(4), 502–513. https://doi.org/10.1109/te.2021.3136914

Zhou, H. (2025). Exploring the dynamic teaching-learning relationship in interactive learning environments. *Interactive Learning Environments*, 1–31. https://doi.org/10.1080/10494820.2025.2462149

Zhou, X., Yang, Q., Bi, L., & Wang, S. (2023). Integrating traditional apprenticeship and modern educational approaches in traditional Chinese medicine education. *Medical Teacher*, *46*(6), 792–807. https://doi.org/10.1080/0142159x.2023.2284661

# APPENDIX

Appendix A: Summary of Key Measures and Results

| Measure | Interactive Mean | Traditional Mean | p-value | Cohen's d | Significant? |
|---|---|---|---|---|---|
| Post-Test | 67.48 | 53.36 | .027 | — | Yes |
| Final Exam | 80.80 | 61.44 | < .001 | — | Yes |
| Behavioral Engagement | 6.48 | 3.88 | < .001 | 1.61 | Yes |
| Emotional Engagement | 5.40 | 3.08 | < .001 | 1.29 | Yes |
| Cognitive Engagement | 2.48 | 4.12 | .001 | -1.05 | Yes (reversed) |
| Extrinsic Motivation | 5.52 | 4.08 | .001 | 1.09 | Yes |
| Intrinsic Motivation | 4.48 | 4.20 | .590 | 0.15 | No |
| General Motivation | 5.12 | 3.36 | .001 | 1.06 | Yes |
| Positive Emotion | 4.00 | 2.40 | < .001 | 1.60 | Yes |
| Negative Emotion | 2.56 | 4.00 | < .001 | -1.38 | Yes |
| Flow | 4.00 | 2.44 | < .001 | 1.59 | Yes |